\documentstyle[aps,epsfig,a4,tighten,twoside,psfig,array]{revtex}

\input epsf

 \newcommand{\zr}[1]{\mbox{\hspace*{#1em}}}
 \newcommand{\ID}{\mbox{{\sf 1}\zr{-0.14}\rule{0.04em}{1.55ex}\zr{0.1}}}

\title{{\noindent\small UNITU-THEP-10/01,  
          FAU-TP3-01/2 \hfill hep-ph/0105142} \\~\\
The Kugo--Ojima Confinement Criterion\\ from
Dyson--Schwinger Equations\thanks{Talk given by R.A.\ at the Workshop on 
{\em Dynamical aspects of the QCD phase transition}, ECT* Trento, 
March 12 - 15, 2001.}} 

\author{Reinhard Alkofer$^a$, Lorenz von Smekal$^b$ and Peter Watson$^a$}
\address{~\\
         $^a$Universit\"at T\"ubingen, Institut f\"ur Theoretische Physik,\\
         Auf der Morgenstelle 14, 72076 T\"ubingen, Germany\\
         E-mail: Reinhard.Alkofer@uni-tuebingen.de\\
         watson@pion01.tphys.physik.uni-tuebingen.de\\ ~\\
         $^b$Universit\"at Erlangen-N\"urnberg,
         Institut f\"ur Theoretische Physik III,\\
         Staudtstr.~7, 91058 Erlangen, Germany\\
         E-mail: smekal@theorie3.physik.uni-erlangen.de}

\begin{document}

\maketitle

\begin{abstract}

Prerequisites of confinement in the covariant and local description of QCD are
reviewed. In particular, the Kugo--Ojima confinement criterion, the positivity
violations of transverse gluon and quark states, and the conditions necessary
to avoid  the decomposition property for colored clusters are discussed. In
Landau gauge QCD, the Kugo--Ojima confinement criterion follows from the 
ghost Dyson--Schwinger equation if the corresponding Green's functions 
can be expanded in an asymptotic series. Furthermore, the infrared behaviour
of the  propagators in Landau gauge QCD as extracted from solutions to
truncated Dyson--Schwinger equations and lattice simulations is discussed
in the light of these issues.  

\end{abstract}

\vskip 5mm

\noindent
{\bf Prerequisites of a Covariant Description of Confinement\\}

The confinement phenomenon in QCD cannot be accommodated within the  standard
framework of quantum field theory. Thereby it is known that covariant quantum
theories of gauge fields require indefinite metric spaces. Maintaining the
much stronger principle of locality, great emphasis has been put on the idea of
relating confinement to the violation of positivity in QCD. Just as in QED,
where the Gupta-Bleuler prescription is to enforce the Lorentz condition on
physical states, a semi-definite {\em physical subspace} can be defined as the
kernel of an operator.   The physical states then correspond to equivalence
classes of states  in this subspace differing by zero norm components.  
Besides transverse photons covariance implies the existence of  longitudinal
and scalar photons in QED. The latter two form metric partners in the
indefinite space. The Lorentz condition eliminates half of these leaving
unpaired states of zero norm which do not contribute to observables. Since the
Lorentz condition commutes with the $S$-Matrix, physical states scatter into
physical ones exclusively. 

Due to the gluon self-interactions the corresponding mechanism is more 
complicated in QCD. Here, the Becchi--Rouet--Stora (BRS) symmetry of the
gauge fixed action proves to be helpful.
Within the framework of BRS algebra, in the simplest version for
the BRS-charge $Q_B$ and the ghost number $Q_c$ given by,
\begin{equation} 
           Q_B^2 = 0 \; , \quad \left[ iQ_c , Q_B \right] = Q_B \; ,
\end{equation}
completeness of the nilpotent BRS-charge $Q_B$ in a state space $\mathcal{V}$
of indefinite metric is assumed. This charge generates the BRS 
transformations by ghost number graded commutators $\{ \, , \}$,
{\it i.e.}, by commutators or anticommutators for fields with even
or odd ghost number, respectively.
The semi-definite subspace 
${\mathcal{V}}_{\mbox{\tiny p}}  = \mbox{Ker}\, Q_B  $
is defined on the basis of this algebra by those states which are annihilated
by the BRS charge $Q_B$. 
Since $Q_B^2 =0 $, this subspace contains the space $ \mbox{Im}\, Q_B $
of so-called daughter
states which are images of others, their parent states in $\mathcal{V}$.
A {\em physical} Hilbert space is then obtained as the 
covariant space of equivalence classes, the BRS-cohomology of states in the
kernel modulo those in the image of $Q_B$,
\begin{equation}
     {\mathcal{H}}(Q_B,{\mathcal{V}}) = {\mbox{Ker}\, Q_B}/{\mbox{Im}\, Q_B} 
       \simeq  {\mathcal{V}}_s \; , 
\end{equation}
which is isomorphic to the space ${\mathcal{V}}_s$ of BRS singlets.    
Completeness is thereby important in the proof of positivity for physical
states \cite{Kug79,Nak90} because it assures the absence of metric
partners of BRS-singlets.

With completeness, all states in $\mathcal{V}$ are either BRS
singlets in ${\mathcal{V}}_s$ or belong to quartets which are 
metric-partner pairs of BRS-doublets (of parent with daughter states).
This exhausts all possibilities. The generalization of the
Gupta--Bleuler condition on physical states, $Q_B |\psi\rangle = 0$ in
$\mathcal{V}_{\mbox{\tiny p}}$, eliminates half of these metric partners
leaving unpaired states of zero norm  which do not contribute to any
observable. This essentially is the quartet mechanism: 
\begin{itemize}
\item[] 
Just as in QED, one such quartet, the elementary quartet, is formed by
the massless asymptotic states of longitudinal and timelike gluons together 
with ghosts and antighosts which are thus all unobservable. 
\item[] 
In contrast to QED, however, one expects the quartet mechanism also 
to apply to transverse gluon and quark states, as far as they exist
asymptotically. A violation of positivity for such states then entails
that they have to be unobservable also. 
\end{itemize}

Asymptotic transverse gluon and quark states
may or may not exist in the indefinite metric space $\mathcal{V}$. If either 
of them do exist and the Kugo--Ojima criterion  (see below) is realized, they
belong to unobservable quartets. In that case, the BRS-transformations of their
asymptotic fields entail that they form these quartets together with
ghost-gluon and/or ghost-quark bound states, respectively \cite{Nak90}.
It is furthermore  crucial for confinement, however, to have a mass gap in
transverse gluon correlations, {\it i.e.}, the massless transverse gluon
states of perturbation theory have to dissappear even though they should
belong to quartets due to color antiscreening \cite{Oeh80,Nis94,Alk00}. 

The interpretation  in terms of transition probabilities  holds between
physical states. For a local operator $A$ to be observable it is necessary to
be BRS-closed, $\{ iQ_B , A \} \, = 0 $, which coincides with the requirement
of its local gauge invariance. It then follows that all states generated
from the vacuum  $|\Omega\rangle$ by any such observable fulfill positivity: 
On the other hand, unobservable, {\it i.e.}, confined, states violate 
positivity.

The remaining dynamical aspect of confinement in this
formulation resides in the cluster decomposition property \cite{Haa96}. 
Including the indefinite metric spaces of covariant gauge
theories it can be summarized as
follows: For the vacuum expectation values of
two local operators $A$ and $B$, translated to a large spacelike
separaration $R$ of each other one obtains the following bounds depending on
the existence of a finite gap $M$ in the spectrum of the 
mass operator in $\mathcal{V}$ \cite{Nak90}   
\begin{eqnarray} 
        \Big|  \langle  \Omega | A(x) B(0) |\Omega \rangle  &-&         
 \langle  \Omega | A(x) |\Omega \rangle  \,  \langle  \Omega
             |  B(0) |\Omega \rangle  \Big|  \\
   && \hskip -.2cm \le  \;  \bigg\{  \begin{array}{ll} 
   \mbox{\small Const.} \, \times \, R^{-3/2 + 2N} \, e^{-MR} \!\!, \quad 
                      & \mbox{mass gap } M \; , \\
   \mbox{\small Const.} \, \times \, R^{-2 + 2N} \,, \;\; 
                      & \mbox{no mass gap} \; ,  \end{array}  \nonumber  
\end{eqnarray}
for $R^2 = - x^2 \to \infty $. Herein, positivity entails that $N = 0$, but a
positive integer $N$ is possible for the indefinite inner product structure in
$\mathcal{V}$. Therefore, in order to avoid the decomposition property
for products of unobservable operators $A$ and $B$ which 
together with the Kugo-Ojima criterion (see below) 
is equivalent to avoiding the
decomposition property for colored clusters, there should 
be no mass gap in the indefinite space $\mathcal{V}$. 
Of course, this implies nothing on the physical spectrum of the mass operator
in $\mathcal{H}$ which certainly should have a mass gap. 
In fact, if the cluster decomposition property holds for a product $A(x) B(0)$ 
forming an observable, it can be shown that both 
$A$ and $B$ are observables themselves. 
This then eliminates the possibility of scattering a physical state into
color singlet states consisting of widely separated colored clusters (the
``behind-the-moon'' problem) \cite{Nak90}. 

Confinement depends on the 
realization of the unfixed global gauge symmetries in this formulation.
The identification of the 
BRS-singlets in the physical Hilbert space $\mathcal{H}$ with
color singlets is possible only if the charge of global gauge transformations
is BRS-exact {\em and} unbroken. The sufficent conditions for this are
provided by the Kugo-Ojima criterion: Considering the 
globally conserved current     
\begin{equation} 
    J^a_\mu = \partial_\nu F_{\mu\nu}^a  + \{ Q_B , D_{\mu}^{ab} \bar c^b \} 
    \qquad (\mbox{with} \; \partial_\mu J^a_\mu = 0 \,) \; ,
       \label{globG}
\end{equation}
one realizes that the first of its two terms corresponds to a coboundary 
with respect to the space-time exterior derivative while the second term 
is a BRS-coboundary with charges denoted by $G^a$ and $N^a$, respectively, 
\begin{equation} 
      Q^a =  \int d^3x \,  \partial_i F_{0 i}^a \,  +\,  \int d^3x \, 
             \{ Q_B , D_{0}^{ab} \bar c^b \} \, = \, G^a \, + \, N^a \; .
        \label{globC}
\end{equation}
For the first term herein there are only two options, it is either ill-defined
due to massless states in the spectrum of $\partial_\nu F_{\mu\nu}^a $, or else
it vanishes. 

In QED massless photon states contribute to the analogues of both currents
in~(\ref{globG}), and both charges on the r.h.s. in (\ref{globC}) are
separately ill-defined. One can employ an arbitrariness in the  definition of
the generator of the global gauge transformations (\ref{globC}), however, to
multiply the first term by a suitable constant so chosen that these massless
contributions cancel. This way one obtains a well-defined and unbroken global
gauge charge which replaces the naive definition in (\ref{globC}) above
\cite{Kug95}. Roughly speaking, there are two independent structures in the
globally conserved gauge currents in QED which both contain massless photon
contributions. These can be combined  to yield one well-defined charge as the
generator of global gauge transformations leaving the other independent
combination (the displacement symmetry) spontaneously broken which lead to the
identification  of photons with massless Goldstone bosons \cite{Nak90,Fer71}. 

If $\partial_\nu F_{\mu\nu}^a $ contains no massless
discrete spectrum on the other hand, {\it i.e.}, if there is no massless
particle pole in the Fourier transform of transverse gluon correlations, then
$G^a \equiv 0$.
In particular, this is the case for channels containing massive vector fields
in theories with Higgs mechanism, and it is expected to be also the case in
any color channel for QCD with confinement for which it actually represents one
of the two conditions formulated by Kugo and Ojima. 
In both these situations one has  
\begin{equation}
                       Q^a \, = \, N^a \, = \, \Big\{   Q_B \, , 
       \int d^3x \,   D_{0}^{ab} \bar c^b \Big\} \; ,
\end{equation}
which is BRS-exact. The second of the two conditions for confinement
is that this charge be well-defined in the whole of the indefinite metric space
$\mathcal{V}$. Together these conditions 
are sufficient to establish that all BRS-singlet physical
states in $\mathcal{H}$ are also color singlets, and that all colored states
are thus subject to the quartet mechanism. The 
second condition thereby provides the essential 
difference between Higgs mechanism and confinement. 
The operator $D_\mu^{ab}\bar c^b$ determining the charge $N^a$ will in
general contain a  {\em massless} contribution from the elementary
quartet due to the asymptotic field $\bar\gamma^a(x)$ in the  
antighost field,  $\bar c^a\, \stackrel{x_0 \to \pm\infty}{\longrightarrow}
\, \bar\gamma^a + \cdots $ (in the weak asymptotic limit), 
\begin{equation}
          D_\mu^{ab}\bar c^b \; \stackrel{x_0 \to \pm\infty}{\longrightarrow}
              \;   ( \delta^{ab} + u^{ab} )\,   \partial_\mu \bar\gamma^b(x) +
                 \cdots  \;  .
\end{equation}
Here, the dynamical parameters $ u^{ab} $ determine the contribution 
of the massless asymptotic state to the composite field $g f^{abc} A^c_\mu
\bar c^b  \, \stackrel{x_0 \to \pm\infty}{\longrightarrow}  \,
u^{ab} \partial_\mu \bar\gamma^b + \cdots $. These parameters can be obtained
in the limit $p^2\to 0$ from the Euclidean correlation functions of this
composite field, {\it e.g.},
\vspace{-.2cm}
\begin{equation}
\int d^4x \; e^{ip(x-y)} \,
\langle  \; D^{ae}_\mu c^e(x) \; gf^{bcd}A_\nu^d(y) \bar c^c (y) \; \rangle
\; =: \; \Big(\delta_{\mu \nu} -{p_\mu p_\nu \over p^2} \Big) \, u^{ab}(p^2)
\; .  \label{Corru}
\end{equation}
The theorem by Kugo and Ojima asserts that all $Q^a = N^a$ do not suffer from
spontaneous breakdown (and are thus well-defined), if and only if
\begin{eqnarray}
                 u^{ab} \equiv u^{ab}(0)  \stackrel{!}{=} - \delta^{ab} \; .
\label{KO1}
\end{eqnarray}
Then the massless states from the elementary quartet do not contribute to 
the spectrum of the current in $N^a$, and the equivalence between physical
BRS-singlet states and color singlets is established.\cite{Kug79,Nak90,Kug95}

In contrast, if $\mbox{det}(  \ID + u ) \not=0$, the global
gauge symmetry generated by the charges $Q^a$ in eq.~(\ref{globC}) is
spontaneuosly broken in each channel in which the gauge potential 
contains an asymptotic massive vector field \cite{Kug79,Nak90}.
While this massive vector state 
results to be a BRS-singlet, the massless Goldstone boson states which 
usually occur in some components of the Higgs field, replace the 
third component of the vector field in the elementary
quartet and are thus unphysical. 
Since the broken charges
are BRS-exact, this
symmetry breaking is not directly observable in the Hilbert space of physical
states $\mathcal{H}$.  

The condition $u = -\ID$ 
Landau gauge be shown by standard arguments employing Dyson--Schwinger
equations and Slavnov--Taylor identities to be 
equivalent to an infrared enhanced ghost propagator \cite{Kug95}.
In momentum space the non-perturbative ghost propagator of Landau gauge QCD  
is related to the form factor occuring in the correlations of
eq.~(\ref{Corru}), 
\begin{equation}
    D_G(p) = \frac{-1}{p^2}      \, \frac{1}{ 1 + u(p^2) } \, , \;\;
                 \mbox{with}  \; \;   
                 u^{ab}(p^2)  = \delta^{ab}  u(p^2) \, . \label{DGdef}
\end{equation}
The Kugo--Ojima confinement criterion, $u(0) = -1$, thus entails that the
Landau gauge ghost propagator should be more singular than a massless particle
pole in the infrared. Indeed, we will present evidence for this exact infrared
enhancement of ghosts in Landau gauge. 

The necessity for the absence of the massless particle pole in $\partial_\nu
F^a_{\mu\nu} $ in the Kugo-Ojima criterion shows that the (unphysical)
massless correlations to avoid the cluster decomposition property are {\em
not} the transverse gluon correlations. An infrared suppressed propagator for
the transverse gluons in Landau gauge confirms this condition. This holds
equally well for the infrared vanishing propagator obtained from
Dyson--Schwinger Equations \cite{Sti96,Sme98} and conjectured in the
studies of the implications of the Gribov horizon \cite{Gri78,Zwa92}, 
as for the infrared suppressed but possibly finite ones extraced from
improved lattice actions for quite large volumes \cite{Bon00}.
The infrared enhanced correlations responsible for the failure of the cluster
decomposition can be identified with the ghost correlations which at
the same time provide for the realization of the Kugo--Ojima criterion in
Landau gauge.

\vskip 5mm


\noindent
{\bf Verifying the Kugo--Ojima Confinement Criterion from the
Dyson--Schwin\-ger Equation for the Ghost Propagator\\}

In Landau gauge the gluon and ghost propagators are parametrized by
the two invariant functions $Z(k^2)$ and $G(k^2)$, respectively
(with $G(k^2) = 1/(1+u(k^2))$, {\it c.f.}, eq.~(\ref{DGdef})). 
In Euclidean momentum space one has
\begin{eqnarray} 
        D_{\mu\nu}(k) = \frac{Z(k^2)}{k^2} \, \left( \delta_{\mu\nu} -
        \frac{k_\mu k_\nu}{k^2} \right)  \; ,\quad
        D_G(k)  &=& - \frac{G(k^2)}{k^2}          \;.
\end{eqnarray}
The non-perturbative infrared behaviour of these 
functions can be studied with employing their  
Dyson--Schwinger equations \cite{Alk00,Rob00}.

The equation for the ghost propagator is the simplest of all QCD
Dyson--Schwinger equations. Besides the ghost and gluon propagators
it contains the ghost-gluon vertex function.  
In Landau gauge this 3-point function needs not to be renormalized.
Furthermore, it becomes bare whenever the out-ghost momentum vanishes.
This has the important consequence that it cannot be singular for 
vanishing ghost momenta.

Furthermore assuming that the QCD Green's  functions can be expanded
in asymptotic series\footnote{Note that this is not possible if the 
infrared slavery picture is correct. An infinite $\beta$-function for vanishing
scales prohibits such an expansion.}, {\it e.g.},
\begin{eqnarray} 
G(p^2;\mu^2) = \sum _{n} d_n \left( \frac {p^2}{\mu^2} \right)^{\delta_n} ,
\end{eqnarray}
the integral in the ghost Dyson--Schwinger equation can be split up in three
pieces. The infrared integral is complicated, and we have not treated it
analytically yet (see, however, ref.\ \cite{Ler01}). The ultraviolet integral,
on the other hand, does not contribute to the infrared behaviour. As a matter
of fact, it is the resulting equation for the  ghost wave function
renormalization constant $\widetilde Z_3$  which allows one to extract definite
information  \cite{Wat01} without using any truncation or specific ansatz
beyond the underlying assumption for the existence of asymptotic infrared
series for QCD Green's functions.

The results are that the infrared behaviour of the gluon and ghost propagators
are uniquely related: The gluon propagator is infrared suppressed as compared
to the one for a free particle, the ghost propagator is infrared enhanced.
This implies that the Kugo--Ojima confinement criterion is satisfied.

\vskip 5mm

\noindent
{\bf A Truncation Scheme for Gluon and Ghost Propagators\\}

\begin{figure}[t]
 \centerline{\epsfxsize=0.7\linewidth \epsfbox{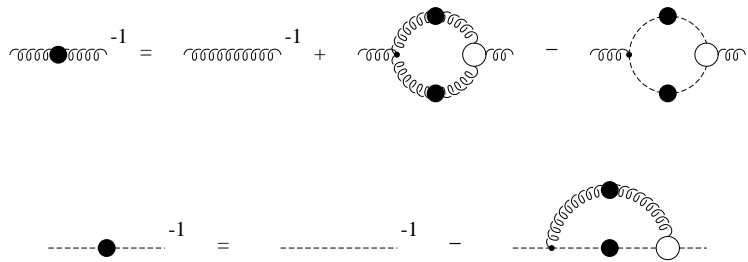}}
\caption{Diagrammatic representation of the truncated system of 
gluon and ghost DSEs.}
\label{GlGh}
\end{figure}

\begin{figure*}[t]

\parbox{.49\linewidth}{\hskip -.1cm\epsfxsize=0.98\linewidth
\epsfbox{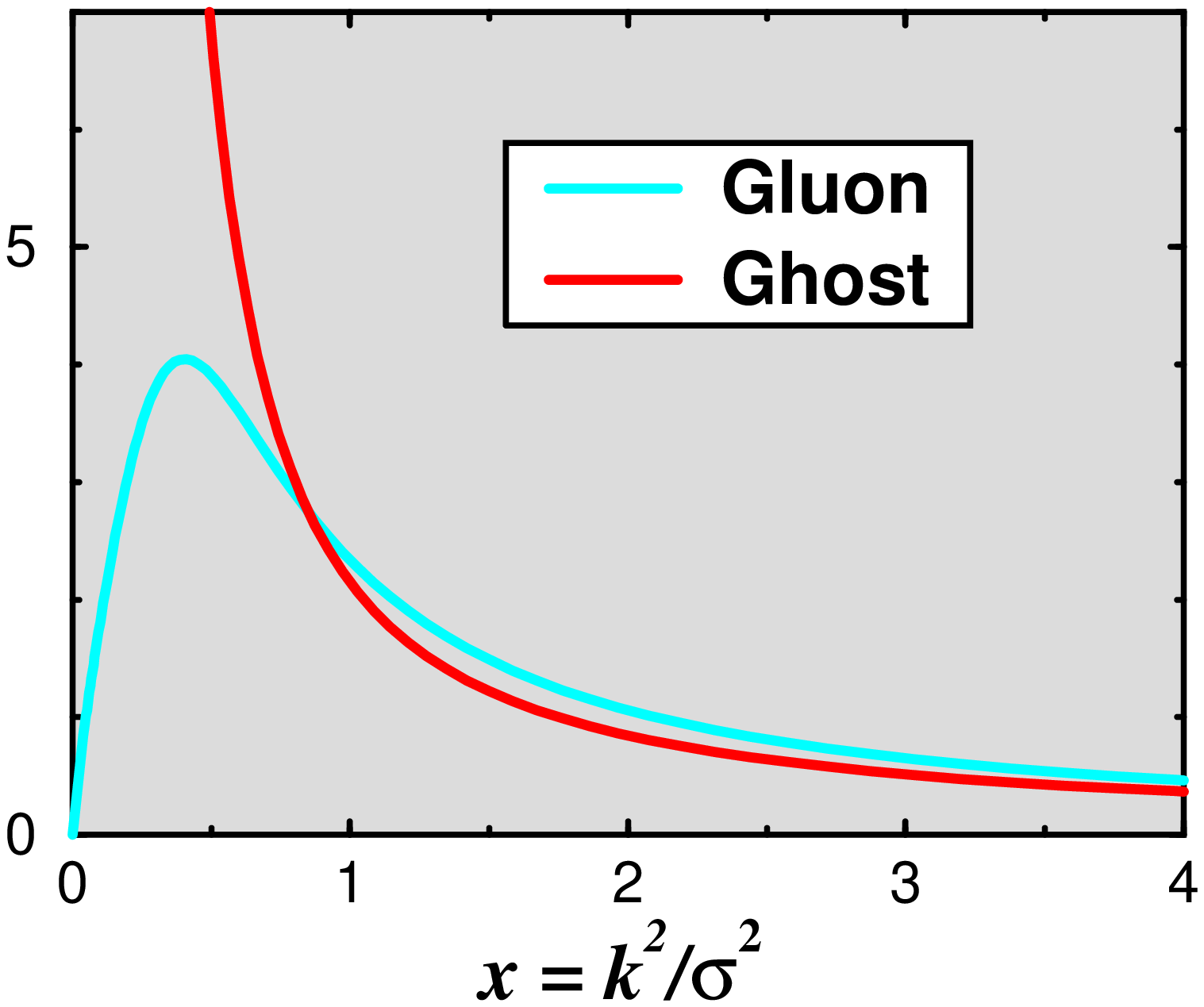}}
\hfill
\parbox{.46\linewidth}{\hskip .5cm\epsfxsize=0.83\linewidth
\epsfbox{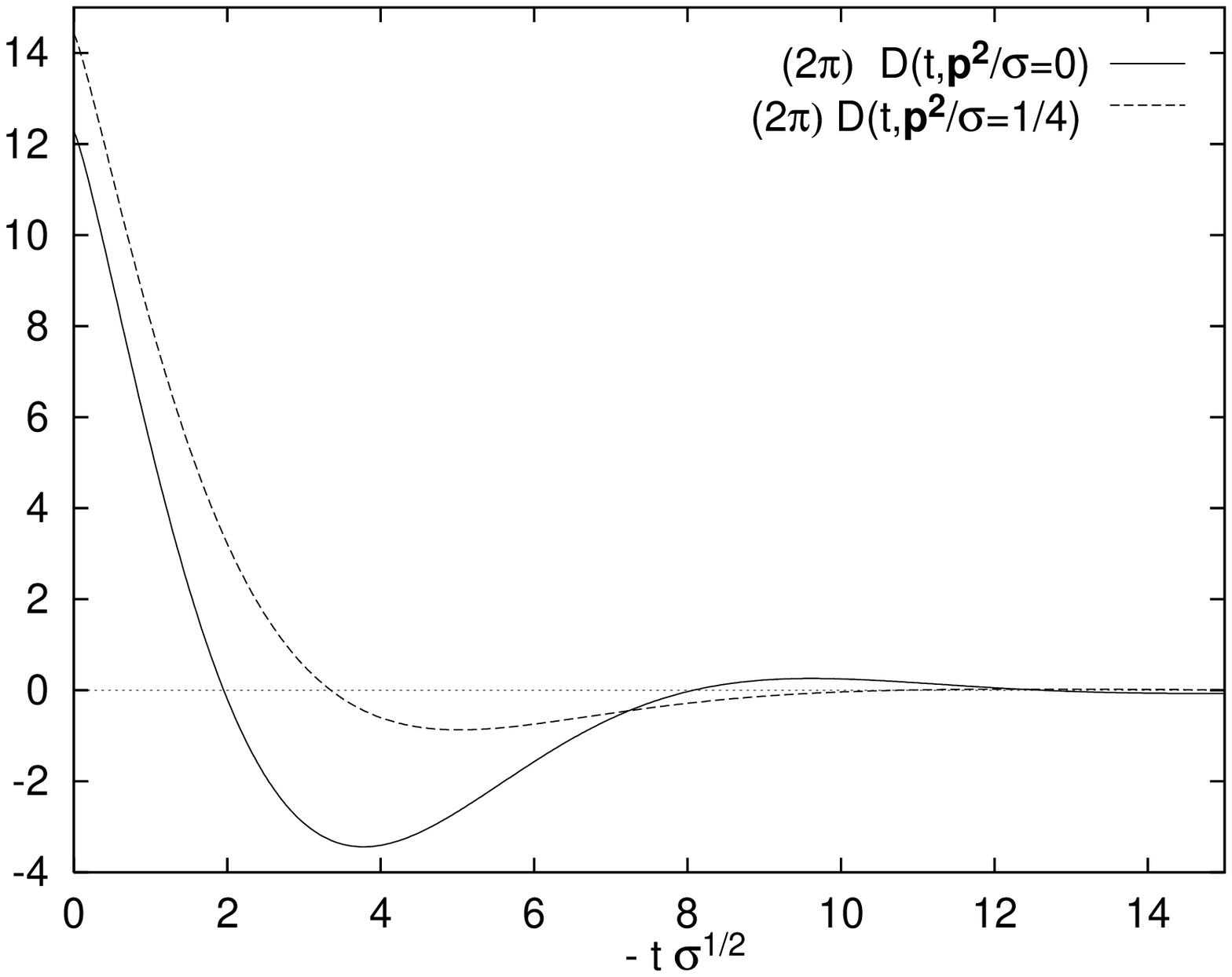}} 

\hskip .1cm\parbox{0.45\linewidth}{\refstepcounter{figure}  \label{ZG}
{\footnotesize Fig. \thefigure:  DSE solutions for $Z(x)$ and
$G(x)$\protect\cite{Sme98}.}}  
\hfill
\parbox{0.4\linewidth}{\refstepcounter{figure} \label{gluon_ft}
{\footnotesize Fig. \thefigure: $D(t, \hbox{\bf p}^2)$ from DSEs for
the gluon\\[-3pt]
renormalization function $Z$ in Fig.~\ref{ZG}.}}
\vskip -.2cm
\end{figure*}

The known structures in the 3-point vertex functions, most importantly from
their Slavnov-Taylor identities and exchange symmtries, have been employed
to establish closed systems of non-linear integral equations that are
complete on the level of the gluon, ghost and quark propagators in Landau
gauge QCD. This is possible with systematically neglecting
contributions from explicit  4-point vertices to the propagator 
Dyson--Schwinger Equations
(DSEs) as well as non-trivial 4-point scattering kernels 
in the constructions of the 3-point vertices \cite{Sme98,Alk00}.
For the pure gauge theory this leads to the propagators DSEs 
diagrammatically represented in Fig.~\ref{GlGh} with
the 3-gluon and ghost-gluon vertices (the open circles) 
expressed in terms of the two functions $Z$ and $G$. 
Employing a one-dimensional approximation one obtains the numerical 
solutions sketched in Fig.~\ref{ZG} \cite{Sme98,Hau98}.

Asymptotic expansions of the solutions in the infrared yield the leading 
infrared behaviour analytically. It is thereby uniquely determined 
by one exponent $\kappa = (61 - \sqrt{1897})/19 \approx 0.92 $,
\begin{equation} 
   Z(k^2) \, \stackrel{k^2\to 0}{\sim}   
      \,  \left(\frac{k^2}{\sigma^2}\right)^{2\kappa}  \quad \mbox{and} \quad
          G(k^2) \, \stackrel{k^2\to 0}{\sim}    
         \, \left(\frac{\sigma^2}{k^2}\right)^{\kappa} \; , \label{IRB}
\end{equation}
for which the bounds $0 < \kappa < 1$ can be established requiring consistency
with Slavnov--Taylor identities \cite{Sme98}.
The renormalization group invariant momentum scale $\sigma $ represents a 
free parameter at this point which is later on fixed by choosing a definite
value for the strong coupling constant at some scale.  
The qualitative infrared behavior in eqs.~(\ref{IRB})
has been also found by studies of further truncated
DSEs \cite{Atk97}. Neither does it thus seem to depend on the particular
3-point vertices nor on employed approximations for angular integrals. 
All these solutions agree qualitatively and confirm the Kugo--Ojima
confinement criterion. 

There are also recent lattice simulations which test this criterion
directly\cite{Nak99}. Instead of $u^{ab} = -\delta^{ab}$ they obtain
numerical values of around $u = -0.7$  for the 
unrenormalized diagonal parts and zero (within  statistical errors) for the
off-diagonal parts. Taking into account the finite size effects on the
lattices employed in the simulations, these preliminary results
might still comply with the Kugo-Ojima confinement criterion.

\vspace{.2cm}
\vspace{.2cm}

Positivity violations of transverse gluon states are manifest in 
the spectral representation of 
the gluon propagator,
\begin{equation} 
       D(p^2) := \frac{Z(p^2)}{p^2}  = \int_0^\infty dm^2
         \, \frac{\rho(m^2)}{p^2 +  m^2} \; .
\end{equation}
>From color antiscreening and unbroken global gauge symmetry
in QCD it follows that the spectral density  
asymptotically is negative and {\em
superconvergent} \cite{Oeh80,Nis94,Alk00}  
\begin{equation} 
  \rho(k^2) \stackrel{k^2\to\infty}{\sim}  - \frac{\gamma g^2}{k^2}
  \Big(g^2 \ln\frac{k^2}{\Lambda^2}\Big)^{-\gamma-1} \hskip -.3cm ,
  \quad\mbox{and}\;
  \int_0^\infty \!\! dm^2  \rho(m^2) = \left(
\frac{g_0^2}{g^2} \right)^\gamma  \to 0 \; , 
\end{equation}
since $ \gamma > 0  $ for $ N_f\le 9 $ in Landau gauge.
This implies that it contains contributions from quartet states 
(and does therefore not need to be gauge invariant unlike in QED).
Here, we consider the one-dimensional Fourier transform 
\begin{eqnarray}
  D(t,\hbox{\bf p}^2) =  \int \frac{dp_0}{2\pi}
    \frac{Z(p_0^2 + \hbox{\bf p}^2)}{p_0^2 + \hbox{\bf p}^2} \; e^{i p_0 t}
               \, = \, 
\int_{\sqrt{\mbox{\footnotesize\bf p}^2}}^\infty  d\omega \, \rho(\omega^2\!
-\! \mbox{\bf p}^2) \,  e^{-\omega t}\; , \label{eq:gluon_FT}
\end{eqnarray}
which for $\rho \ge 0$ had to be positive definite (and one had
$\frac{d^2}{dt^2} \ln D(t,\hbox{\bf p}) \ge 0$).
This is clearly not the case for the DSE solution shown in
Fig.~\ref{gluon_ft} which violates reflection positivity \cite{Alk00,Sme98}.
Even though no negative $D(t, \hbox{\bf p}^2)$ have been reported in 
lattice calculations yet, the available results \cite{Man99} 
agree in indicating that ln $D(t, \hbox{\bf p}^2)$ is
not the convex function of the Euclidean time it should be for positive
$\rho$ \cite{Man87,Nak95}. These are non-perturbative verifications of the
positivity violation for transverse gluon states which already occur in
perturbation theory. More significant for
confinement is the fact that no massless single transverse gluon 
contribution to $\rho$ exists for $Z(0) = 0$.

\begin{figure*}[t]
\parbox{.49\linewidth}{\hskip -.5cm\epsfxsize=1.07\linewidth
\epsfbox{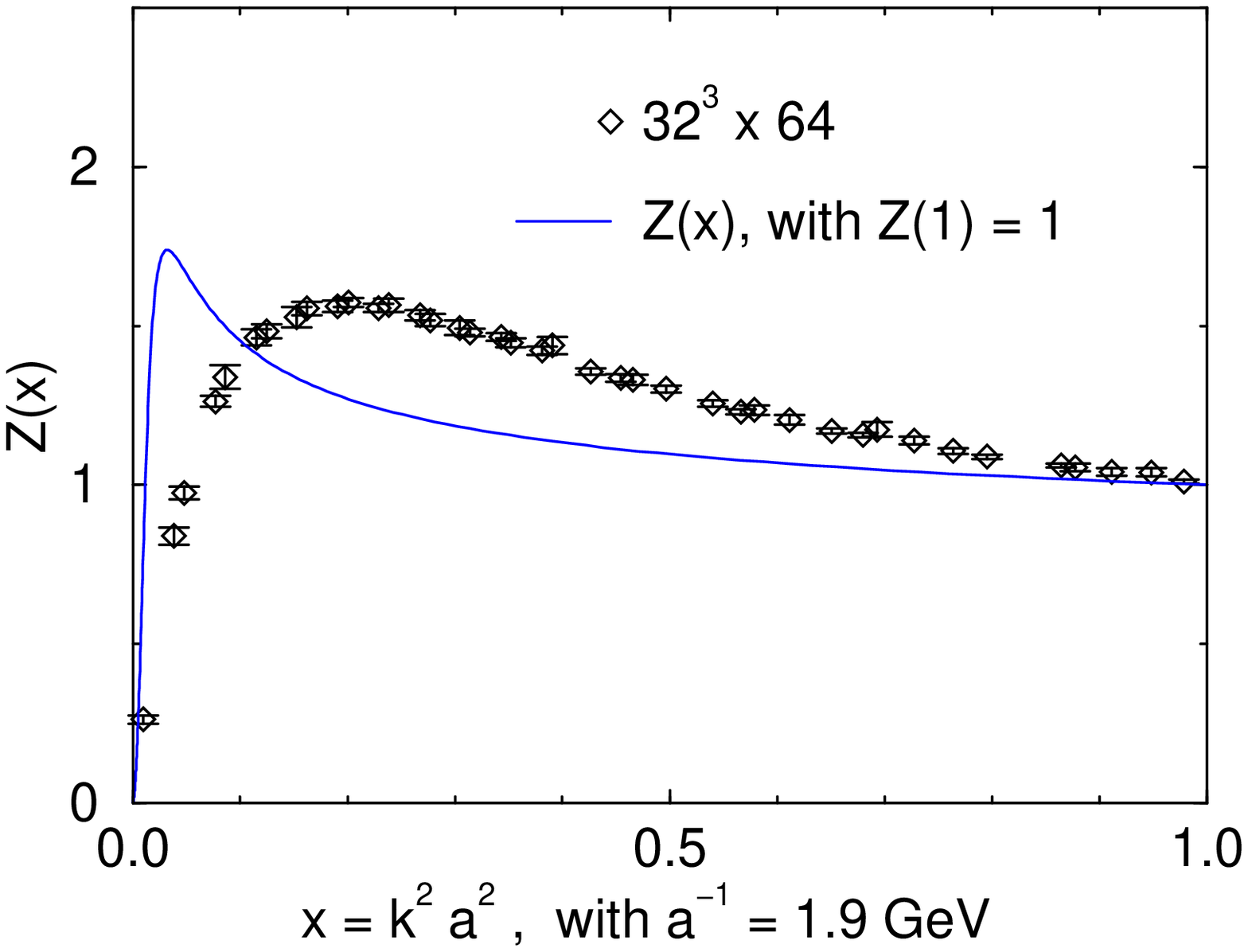}}
\hskip -.1cm
\parbox{.48\linewidth}{\hskip .2cm
\epsfxsize=0.97\linewidth\epsfbox{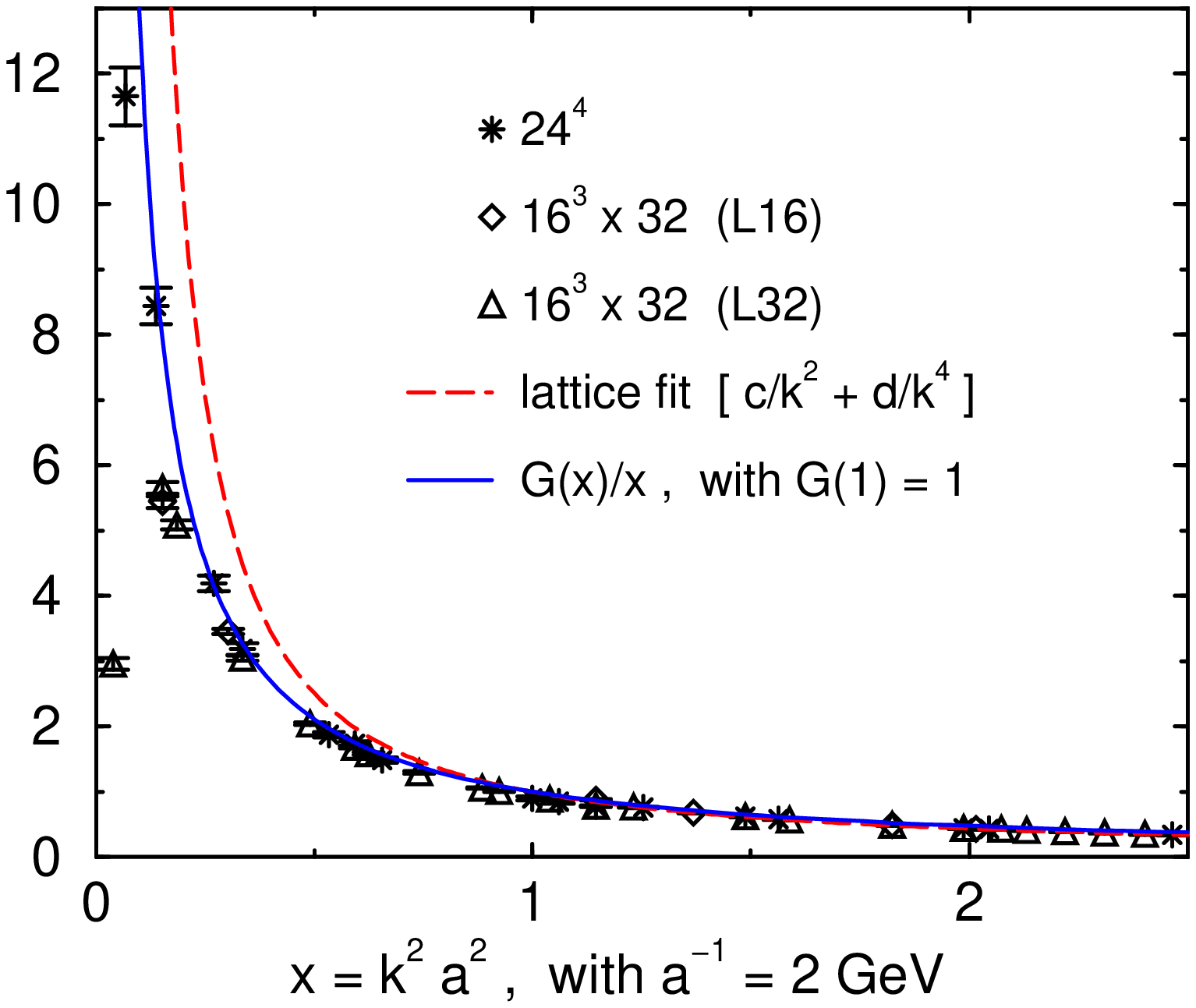}}

\hskip .2cm\parbox{0.42\linewidth}{
\refstepcounter{figure} \label{Der98} 
{\footnotesize Fig. \thefigure: The gluon renormalization function\\[-3pt] 
from the DSE solutions of Ref.~\protect\cite{Sme98} (solid line)\\[-3pt] 
and from the lattice data of Ref.~\protect\cite{Lei98}.}}  
\hskip .3cm \hfill
\parbox{0.48\linewidth}{\refstepcounter{figure}  \label{Sum95col} 
{\footnotesize Fig. \thefigure: The ghost propagator from
DSEs in Ref.~\protect\cite{Sme98}\\[-3pt] 
(solid line) compared to data and fit (dashed with\\[-3pt]  
$ca^2 =0.744 , \; da^4=0.256 $ for $x \ge 2$) from Ref.~\protect\cite{Sum96}.}}
\end{figure*}

Confirmation of the important result that the gluon renormalization function
vanishes in the infrared and no massless asymptotic transverse gluon states
occur, {\it i.e.}, $Z(0) =0$, is seen in Fig.~\ref{Der98}, where 
the DSE solution of Fig.~\ref{ZG} is compared to lattice data \cite{Lei98} 
and it was further verified recently with improved lattice actions for large
volumes \cite{Bon00}. This infrared suppression as seen in lattice
calculations thereby seems to be weaker than the DSE result, 
apparently giving rise to an infrared finite gluon
propagator $D(k) \sim Z(k^2)/k^2 $ (corresponding to an exponent 
$\kappa = 1/2 $ in~(\ref{IRB})), but a vanishing one does not seem to be 
ruled out for the infinite volume limit \cite{Cuc97}.
Similar results with finite $D(0)$ are also 
reported from the Laplacian gauge which practically 
avoids Gribov copies \cite{Ale00}.

\vspace{.2cm}
\vspace{.2cm}

The infrared enhanced DSE solution for ghost propagator is compared
to lattice data in Fig.~\ref{Sum95col}. 
One observes quite compelling agreement, the numerical DSE solution fits the
lattice data at low momenta ($x \le 1$) significantly better than the fit to an
infrared singular form with integer exponents, $D_G(k^2) = c/k^2  + d/k^4$.
Though low momenta ($x<2$) were excluded in this fit, the authors concluded
that no reasonable fit of such a form was otherwise possible \cite{Sum96}.
Therefore, apart from the question about a possible maximum at the very
lowest momenta, the lattice calculation seems to confirm the infrared
enhanced ghost propagator 
with a non-integer exponent $0 < \kappa < 1$. 
The same qualitative conclusion has in fact been obtained in a more recent
lattice calculation of the ghost propagator in $SU(2)$ \cite{Cuc97}, where its
infrared dominant part was fitted best by $D_G \sim 1/(k^2)^{1+\kappa}$ for
an exponent $\kappa $ of roughly $ 0.3$ (for $\beta = 2.7$). 

To summarize, the qualitative infrared behavior in eqs.~(\ref{IRB}), an
infrared suppression of the gluon propagator together with an infrared
enhanced ghost propagator as predicted by the Kugo-Ojima criterion for the
Landau gauge, is confirmed by the presently availabe lattice
calculations. The exponents obtained therein ($0.3 < \kappa \le 0.5 $) both
seem to be consistently smaller than the one obtained in solving their DSEs.
Whether also the lattice data is thereby determined 
by one unique exponent $0<\kappa <1 $ for 
the infrared behavior of both propagators, has not yet been investigated
to our knowledge.
An independent confirmation of this combined infrared behavior which is 
indicative of an infrared fixed point would support the existence of the
unphysical massless states that are necessary to circumvent the decomposition
property for colored clusters.

\bigskip

\noindent
{\bf Acknowledgements}

\noindent
R.\ A.\  thanks Sebastian Schmidt and David Blaschke for organizing 
this stimulating workshop. 

\noindent
This work has been supported in part by the DFG under contract Al 279/3-3
and under contract We 1254/4-2.

\end{document}